\documentclass[twocolumn,showpacs,nofootinbib,preprintnumbers,amsmath,amssymb]{revtex4}

\usepackage[dvips]{graphicx}

\newcommand{\lsim}{\mathrel{\vcenter{\hbox{$<$}\nointerlineskip\hbox{$\sim$}}}}
\newcommand{\gsim}{\mathrel{\vcenter{\hbox{$>$}\nointerlineskip\hbox{$\sim$}}}}

\begin{document}
\title{Oscillating Quintessence}

\author{Je-An~Gu}
\email{jagu@phys.cts.nthu.edu.tw} %
\affiliation{National Center for Theoretical Sciences, Hsinchu 30013, Taiwan, R.O.C.} %

\begin{abstract}
An oscillating scalar field as a quintessence model for dark
energy is proposed. The case of a power-law potential is
particularly intriguing and is the focus of the present article.
In this model the equation of state $w_\textsc{oq}$ of dark energy
is a constant determined simply by the power $n$ in the potential
through $w_\textsc{oq} = (n-2)/(n+2)$. Accordingly, when $0<n<1$,
the oscillating quintessence can provide repulsive gravity and
drive the cosmic acceleration. The condition for oscillation and
the constraints from observations are investigated. For this new
scenario a specific natural model with much less fine tuning is
presented.
\end{abstract}


\maketitle


\textit{Introduction.} The existence of the accelerated expansion
of the present universe was discovered in 1998 via type Ia
supernova (SN Ia) distance measurement
\cite{Perlmutter:1999np,Riess:1998cb} and
further reinforced by recent observations %
\cite{SNLS,Riess:2006fw,WMAP,Tegmark:2006}.
This mysterious phenomenon is one of the most important unsolved
problems 
in this century. Since this discovery a variety of models have
been proposed and many of them so far survive under continuingly
updated observational results. These models involve diverse
approaches and strategies, including (1) introducing new energy
contents which provide negative pressure and repulsive gravity,
e.g.\ a positive cosmological constant
\cite{Krauss:1995yb,Ostriker:1995su,Liddle:1996pd} and a dynamical
scalar field such as quintessence
\cite{Caldwell:1998ii,Gu:2001tr,Boyle:2001du}, phantom
\cite{Caldwell:1999ew}, etc., as generally called dark energy, (2)
introducing new ingredients in the geometry/gravity part, such as
modifying the gravity action
\cite{Carroll:2003wy,Lue:2003ky,Arkani-Hamed:2003uy} and invoking
extra dimensions
\cite{Deffayet:2001pu,Gu:2001ni,Gu:2002mz,Gu:2004fg,Aghababaie:2003wz,Burgess:2004kd,CasimirDE:2007},
and (3) taking into consideration the effects of inhomogeneities
of the universe (e.g.\ see \cite{Chuang:2005yi,Buchert:2007ik} and
references therein). Among these models, quintessence will be the
focus in the present article.

For a field (or a physical object) and its possible evolution
patterns, the oscillation-like behavior (including orbiting
around) is the most familiar because it has been seen frequently
in nature from particle physics of subatomic scales, daily lives
of human scales, to our universe of astronomical scales.
Oscillation is therefore the most natural mode for a field to
consider. Nevertheless, with regard to the quintessence as a
realization of dark energy with negative pressure and repulsive
gravity, it is straightforward to think that oscillating modes
correspond to excitations or particles that provide nonnegative
pressure and attractive gravity and therefore can hardly do the
job. For example \cite{Gu:2001}, for $V=\frac{1}{2} m^2 \phi^2$,
the oscillating scalar field is a linear combination of sin- and
cosine-wave modes, $\cos(k \cdot x + \theta_0)$, while each mode
provides nonnegative pressure with the equation of state ranging
from $0$ to $1/3$, corresponding to the wave number $|\vec{k}|$
from the small ($\ll m$) to the large ($\gg m$).\footnote{The
behaviors of the small- and the large-wave-number modes are
consistent with those of the non-relativistic and the relativistic
particles, respectively \cite{Gu:2001}.} If insisting to utilize
this potential to construct the quintessence model, one requires
extremely small mass $m \lsim H_0 \sim 10^{-33}\textrm{eV}$ and
extremely large amplitude $|\phi| \gsim M_\textrm{pl} \sim
10^{28}\textrm{eV}$ to accommodate a slowly (enough) evolving mode
(instead of oscillation) for dark energy. Accordingly, most of the
efforts at the quintessence model construction are devoted to
slowly rolling evolution patterns under a rather smooth and flat
potential that dominates over the kinetic energy of the
quintessence at all times in the recent epoch. For example, in the
tracker quintessence model \cite{Zlatev:1999tr,Steinhardt:1999nw}
the potential goes to zero asymptotically when the scalar field
goes to infinity along with time, as a realization of the running
away evolution pattern.

On the contrary, in the present article, a new quintessential dark
energy scenario involving an oscillating scalar field is proposed.
In particular, the present article is focused on the power-law
potential $V(\phi) = M^{4-n} |\phi|^{n}$ ($n>0,M>0$). In this
model, as going to be presented, the equation of state of dark
energy is a constant determined simply by the power in the
potential:
\begin{equation} \label{eq:w-oq}
w_\textsc{oq} = \frac{n-2}{n+2} \, ,
\end{equation}
where the subscript ``\textsc{oq}'' stands for oscillating
quintessence. As a result, when $0<n<1$, we have $-1<
w_\textsc{oq} < -1/3$ and accordingly the oscillating quintessence
can drive the cosmic acceleration.


\vspace{0.5em}

\textit{Analysis.} Consider a Friedmann-Lemaitre-Robertson-Walker
(FLRW) universe which is described by the Robertson-Walker metric,
\begin{equation}
ds^2 = dt^2 - a^2(t) \left( \frac{dr^2}{1-kr^2} + r^2 d \Omega ^2
\right) \, , %
\label{eq:RW metric}
\end{equation}
and at the present epoch is dominated by pressureless matter and
quintessence. The quintessence is represented by a scalar field
with the Lagrangian density:
\begin{equation} \label{eq:scalar Lagrangian}
\mathcal{L} = \sqrt{|g|} \left[ \frac{1}{2} g^{\mu \nu } (\partial
_{\mu} \phi ) (\partial _{\nu} \phi ) - V(\phi ) \right] \, .
\end{equation}
For the evolution of this universe, the governing equations are as
follows.
\begin{equation}
\left( \frac{\dot{a}}{a} \right) ^2 + \frac{k}{a^2} = \frac{8 \pi
G}{3} \rho = \frac{8 \pi
G}{3} \left( \rho_\textrm{m} + \rho_{\phi} \right) \, , %
\label{eq:Friedmann eqn}
\end{equation}
\begin{equation}
\frac{\ddot{a}}{a} = - \frac{4 \pi G}{3} \left( \rho + 3p \right)
= - \frac{4 \pi G}{3} \left( \rho_\textrm{m} + \rho_{\phi} + 3p_{\phi} \right) \, , %
\label{eq:accel eqn}
\end{equation}
\begin{equation}
\frac{\partial^2 \phi}{\partial t^2} + 3H \frac{\partial
\phi}{\partial t} - \frac{1}{a^2}\nabla ^2 \phi + V'(\phi ) =0 \; ,%
\label{eq:phi field eqn}
\end{equation}
where the Hubble expansion rate $H \equiv \dot{a}/a$ and the
energy density and pressure of the quintessence are given by
\begin{eqnarray}
\rho_{\phi} &=& \frac{1}{2} \left( \frac{\partial \phi}{\partial
                 t} \right) ^2 + \frac{1}{2a^2} \left(
                 \nabla \phi \right) ^2 + V(\phi ) \; , \label{phi rho} \\
p_{\phi}    &=&  \frac{1}{2} \left( \frac{\partial \phi}{\partial
               t} \right) ^2 - \frac{1}{6a^2} \left(
              \nabla \phi \right) ^2 - V(\phi ) \; . \label{phi p}
\end{eqnarray}
For simplicity, in the rest of the present article, the spatial
curvature and the spatial derivative terms will be ignored
respectively by the assumption that our universe is flat spatially
($k=0$) and the spatial dependence of the quintessence is very
weak.

In the present article the following power-law potential is
considered.
\begin{equation} \label{eq:power-law potential}
V(\phi) = M^{4-n} \left| \phi \right|^{n} \, , \quad n>0 \, , M>0
\, .
\end{equation}
For $0 < n \leq 1$, in order to regularize the potential around
$\phi=0$, the potential is modified as follows.
\begin{equation} \label{eq:regularized power-law potential}
V(\phi) = \frac{M^{4-n} |\phi|^{1+n}}{|\phi|+\epsilon} \, , \quad
\epsilon>0 \, .
\end{equation}
This modification can be ignored when $|\phi| \gg \epsilon$.

Here we consider an oscillating mode of the quintessence whose
period $T$ is much smaller than the Hubble time, i.e., $T \ll
H^{-1}$, and study the quantities averaged over a period, $\langle
\mathcal{O} \rangle$. From the field equation (\ref{eq:phi field
eqn}) where the damping term $3H\dot{\phi}$ can be ignored for $T
\ll H^{-1}$, it is straightforward to obtain the relation between
the kinetic and the potential contributions to the averaged
quintessence energy density:\footnote{The derivation is similar to
that of the virial theorem.}
\begin{equation} \label{eq:K-V relation}
\langle K \rangle = \frac{n}{2} \langle V \rangle \, ,
\end{equation}
where $K \equiv \dot{\phi}^2/2$ and the over-head dot denotes the
time derivative. Thus, for an oscillating quintessence under a
power-law potential, the equation of state (averaged over a period
$T$ or a time scale much larger than the period) is a constant
determined simply by the power $n$, as given in Eq.\
(\ref{eq:w-oq}):
\begin{equation}
w_\textsc{oq} = \frac{n-2}{n+2} \, . \nonumber
\end{equation}

Regarding different $w_\textsc{oq}$ corresponding to different
power $n$, several cases are listed in the following table.
\begin{center}
\begin{tabular}{|c|rrrrrccc|} \hline
            $n$ & $0^{+}$  &  $0.01$ &  $0.05$ &  $0.1$ &  $0.5$ &    $1$ & $2$ &   $4$ \\ %
\hline
$w_\textsc{oq}$ & $-1^{+}$ & $-0.99$ & $-0.95$ & $-0.9$ & $-0.6$ & $-1/3$ & $\;0\;$ & $1/3$ \\
\hline
\end{tabular}
\end{center}
In particular, when $0<n<1$, the oscillating quintessence provides
negative pressure and repulsive gravity with the equation of state
between $-1$ and $-1/3$, and therefore has the ability to drive
the cosmic expansion to accelerate.

In addition to the equation of state, the other two essential
quantities are the energy density $\rho_\textsc{oq}$ and the
period $T$ of the oscillating quintessence:
\begin{eqnarray} \label{eq:rho-oq}
\rho_\textsc{oq} = \langle \rho_{\phi} \rangle &=& \langle K
\rangle + \langle V \rangle = \left(1+\frac{n}{2}\right) \langle V
\rangle \nonumber
\\
&=& V_\textrm{max} = M^{4-n} |\phi|_\textrm{max}^n \, ,
\end{eqnarray}
where the subscript ``max'' denotes the maximal value within a
period $T$ and $|\phi|_\textrm{max}$ stands for the amplitude of
the oscillation;
\begin{equation} \label{eq:T}
T = \frac{\sqrt{8\pi}}{n}
\frac{\Gamma(\frac{1}{n})}{\Gamma(\frac{1}{2}+\frac{1}{n})} \cdot
M^{1-4/n} \rho_\textsc{oq}^{1/n-1/2} ,
\end{equation}
or
\begin{eqnarray} \label{eq:HT}
H T &=& \frac{8\pi}{\sqrt{3}n}
\frac{\Gamma(\frac{1}{n})}{\Gamma(\frac{1}{2}+\frac{1}{n})} \cdot
M_\textrm{pl}^{-1} M^{1-4/n} \rho^{1/2} \rho_\textsc{oq}^{1/n-1/2}
\nonumber \\
&=& \frac{8\pi}{\sqrt{3}n}
\frac{\Gamma(\frac{1}{n})}{\Gamma(\frac{1}{2}+\frac{1}{n})} \cdot
\left( \frac{\rho_\textsc{oq}}{\rho} \right)^{-\frac{1}{2}} %
\left( \frac{\rho_\textsc{oq}^{1/4}}{M_\textrm{pl}} \right) %
\left( \frac{\rho_\textsc{oq}^{1/4}}{M} \right)^{\frac{4}{n}-1} \nonumber \\
&=& \frac{8\pi}{\sqrt{3}n}
\frac{\Gamma(\frac{1}{n})}{\Gamma(\frac{1}{2}+\frac{1}{n})} \cdot
\sqrt{\frac{\rho}{\rho_\textsc{oq}}} %
\left( \frac{|\phi|_\textrm{max}}{M_\textrm{pl}} \right) \, ,
\end{eqnarray}
where the Planck scale $M_\textrm{pl} \equiv G_N^{-1/2} \cong 1.2
\times 10^{19}\,$GeV and the total energy density $\rho$ in unit
of the current value [$\rho(0)=\rho_\textrm{c}$] as a function of
the redshift $z$ is as follows.
\begin{eqnarray}
\hspace*{-2em} %
\lefteqn{\frac{\rho(z)}{\rho(0)} = \frac{\rho(z)}{\rho_\textrm{c}}} \nonumber \\
\hspace*{-2em} &=& \Omega_\textsc{oq} (1+z)^{6n/(2+n)} +
\Omega_\textrm{m} (1+z)^{3} + \Omega_{\gamma} (1+z)^{4} , %
\label{eq:rho(z) in unit of rhoc}
\end{eqnarray}
if only the contributions from quintessence, matter and photons
are considered. %
For the present time we have
\begin{eqnarray} \label{eq:H0T0}
H_0 T_0 &=& \frac{\sqrt{3}}{n}
\frac{\Gamma(\frac{1}{n})}{\Gamma(\frac{1}{2}+\frac{1}{n})} %
\left( \frac{3}{8\pi} \right)^{\frac{4}{n}-1} \nonumber \\
&& \hspace{2em} \cdot \; %
\Omega_\textsc{oq}^{\frac{1}{n}-\frac{1}{2}} %
\left( \frac{M}{M_\textrm{pl}} \right)^{1-\frac{2}{n}} %
\left( \frac{M}{H_0} \right)^{-\frac{2}{n}} \nonumber  \\
&=& \frac{8\pi}{\sqrt{3}n}
\frac{\Gamma(\frac{1}{n})}{\Gamma(\frac{1}{2}+\frac{1}{n})} \cdot
\Omega_\textsc{oq}^{-1/2} %
\left( \frac{|\phi|_\textrm{max}(0)}{M_\textrm{pl}} \right) ,
\end{eqnarray}
where ``$0$'' denotes the present time.

%

\vspace{0.5em}

\textit{Requirements and Constraints.} According to the
observational results, we have the following information as our
input \cite{Yao:2006px}: $H_0 \cong 73 \, \textrm{km} \,
\textrm{s}^{-1} \, \textrm{Mpc}^{-1}$, $\rho_\textrm{c} \cong 4.3
\times 10^{-11}\,$eV$^4$, $\Omega_\textrm{m} \cong 0.24$,
$\Omega_{\gamma} \cong 4.6 \times 10^{-5}$, the dark energy
(``x'') density fraction $\Omega_\textrm{x} \cong 0.75$ and the
equation of state $w_\textrm{x} < -0.9$ (1$\sigma$) (for the case
of a constant $w_\textrm{x}$). For the oscillating quintessence to
play the role of dark energy, from Eqs.\ (\ref{eq:w-oq}) and
(\ref{eq:rho-oq}) the constraints $w_\textrm{x} < -0.9$
(1$\sigma$) and $\Omega_\textrm{x} \cong 0.75$ respectively
requires
\begin{eqnarray}
0 < &n& < 0.1 \, , \label{eq:constraint on n} \\
V_\textrm{max}(0) &=& M^{4-n} |\phi|_\textrm{max}^n (0) \nonumber \\ %
&\cong& 0.75 \, \rho_\textrm{c} \cong 3.2 \times 10^{-11} \,
\textrm{eV}^4 \, ,
\end{eqnarray}
as two conditions for the three parameters, $n$, $M$ and
$|\phi|_\textrm{max} (0)$, in the power-law potential.

The condition for the oscillation is
\begin{equation}
H T \ll 1 \quad \textrm{from the present back to some early time.}
\end{equation}
The condition at the present time ($H_0 T_0 \ll 1$) requires
\begin{equation} \label{eq:constraint on M from H0T0}
M \gg \left[ \frac{8\pi}{\sqrt{3}n}
\frac{\Gamma(\frac{1}{n})}{\Gamma(\frac{1}{2}+\frac{1}{n})} \cdot
\Omega_\textsc{oq}^{1/n-1/2} \rho_\textrm{c}^{1/n}
M_\textrm{pl}^{-1} \right]^{\frac{n}{4-n}} \, ,
\end{equation}
or, equivalently,
\begin{equation} \label{eq:constraint on phiMax from H0T0}
|\phi|_\textrm{max}(0) \ll \frac{8\pi}{\sqrt{3}n}
\frac{\Gamma(\frac{1}{n})}{\Gamma(\frac{1}{2}+\frac{1}{n})} \cdot
\sqrt{\Omega_\textsc{oq}} M_\textrm{pl} \, ,
\end{equation}
i.e., the current amplitude of the oscillation
$|\phi|_\textrm{max}(0)$ should be smaller than or roughly around
the same scale of the Planck scale $M_\textrm{pl}$.

In addition, the condition $HT \ll 1$ gives a lower limit to the
time, i.e.\ an upper limit to the redshift $z$, with respect to
the beginning of the oscillation. From Eqs.\
(\ref{eq:HT})--(\ref{eq:H0T0}), we have
\begin{eqnarray}
\frac{H T}{H_0 T_0} &=& (1+z)^{3(2-n)/(2+n)}
\sqrt{\rho(z)/\rho(0)}
\nonumber \\
&\equiv& F_n(z) \, ,
\end{eqnarray}
where the ratio of the total energy density, $\rho(z)/\rho(0)$, is
given in Eq.\ (\ref{eq:rho(z) in unit of rhoc}). Note that, for
$0<n<1$, $F_n(z)$ is a monotonically increasing function of $z$.
Accordingly, $HT \ll 1$ requires
\begin{equation}
z \ll z_\textrm{max} \equiv F_n^{-1} \left( \frac{1}{H_0 T_0}
\right) \, .
\end{equation}

Several examples with various values of $n$ and $M$ fulfilling the
above constraints and requirements are listed in Table
\ref{table:values for n M etc}. Among these examples the case with
$n \simeq 0.1$ and $M = \rho_\textrm{c}^{1/4}$ is particularly
interesting. In this case both the two free dimensionful
parameters, $M$ and $|\phi|_\textrm{max}(0)$, are on the same
scale of $\rho_\textrm{c}^{1/4}$, i.e., the energy scale of the
present universe regarding the energy density. In addition, the
period of the oscillation at the present time $T_0$ is also
roughly around the same scale, $\rho_\textrm{c}^{-1/4}$, and the
oscillation started around $z = 7 \times 10^6$ when the
temperature of the universe was around $10^7$K (i.e.\ $1$keV).


\begin{table}[h!] \caption{Examples with various values of $n$ and $M$ for demonstration.}
\begin{tabular}{|cccccc|}
\multicolumn{6}{l}{(1) $n=0.5$, $w_\textsc{oq}=-0.6$, %
$M \gg 1.6 \; 10^{-7}\textrm{eV}$, %
$\frac{|\phi|_\textrm{max}(0)}{M_\textrm{pl}} \ll 19$} \\
\hline %
$M$ &
\begin{tabular}{c}
($\rho_\textrm{c}$) \\
$2.56 \; 10^{-3}\textrm{eV}$
\end{tabular}
& eV & MeV & TeV &
\begin{tabular}{c}
($M_\textrm{pl}$) \\
$10^{28}\textrm{eV}$
\end{tabular}
\\ %
\hline %
$\log \frac{|\phi|_\textrm{max}(0)}{M_\textrm{pl}}$ %
& $-31$ & $-49$ & $-91$ & $-133$ & $-245$ %
\\
$\log H_0 T_0$ %
& $-30$ & $-48$ & $-90$ & $-132$ & $-244$ %
\\
$\log z_\textrm{max}$ %
& $8.3$ & $13$ & $24$ & $35$ & $65$ %
%
\\ \hline
%
\multicolumn{6}{l}{(2) $n=0.1$, $w_\textsc{oq}=-0.9$, %
$M \gg 4.3 \; 10^{-4}\textrm{eV}$, %
$\frac{|\phi|_\textrm{max}(0)}{M_\textrm{pl}} \ll 40$} \\
\hline %
$M$ &
\begin{tabular}{c}
($\rho_\textrm{c}$) \\
$2.56 \; 10^{-3}\textrm{eV}$
\end{tabular}
& eV & MeV & TeV &
\begin{tabular}{c}
($M_\textrm{pl}$) \\
$10^{28}\textrm{eV}$
\end{tabular}
\\ %
\hline %
$\log \frac{|\phi|_\textrm{max}(0)}{M_\textrm{pl}}$ %
& $-32$ & $-133$ & $-367$ & $-601$ & $-1225$ %
\\
$\log H_0 T_0$ %
& $-30$ & $-131$ & $-365$ & $-599$ & $-1223$ %
\\
$\log z_\textrm{max}$ %
& $6.9$ & $28$ & $78$ & $128$ & $260$ %
%
\\ \hline
%
\multicolumn{6}{l}{(3) $n=0.05$, $w_\textsc{oq}=-0.95$, %
$M \gg 10^{-3}\textrm{eV}$, %
$\frac{|\phi|_\textrm{max}(0)}{M_\textrm{pl}} \ll 57$} \\
\hline %
$M$ &
\begin{tabular}{c}
($\rho_\textrm{c}$) \\
$2.56 \; 10^{-3}\textrm{eV}$
\end{tabular}
& eV & MeV & TeV &
\begin{tabular}{c}
($M_\textrm{pl}$) \\
$10^{28}\textrm{eV}$
\end{tabular}
\\
\hline %
$\log \frac{|\phi|_\textrm{max}(0)}{M_\textrm{pl}}$ %
& $-33$ & $-238$ & $-712$ & $-1186$ & $-2450$ %
\\
$\log H_0 T_0$ %
& $-31$ & $-236$ & $-710$ & $-1184$ & $-2448$ %
\\
$\log z_\textrm{max}$ %
& $6.9$ & $49$ & $147$ & $244$ & $505$ %
%
\\ \hline
%
\multicolumn{6}{l}{(4) $n=0.01$, $w_\textsc{oq}=-0.99$, %
$M \gg 2 \; 10^{-3}\textrm{eV}$, %
$\frac{|\phi|_\textrm{max}(0)}{M_\textrm{pl}} \ll 126$} \\
\hline %
$M$ &
\begin{tabular}{c}
($\rho_\textrm{c}$) \\
$2.56 \; 10^{-3}\textrm{eV}$
\end{tabular}
& eV & MeV & TeV &
\begin{tabular}{c}
($M_\textrm{pl}$) \\
$10^{28}\textrm{eV}$
\end{tabular}
\\
\hline %
$\log \frac{|\phi|_\textrm{max}(0)}{M_\textrm{pl}}$ %
& $-43$ & $-1077$ & $-3471$ & $-5865$ & $\sim -10^4$ %
\\
$\log H_0 T_0$ %
& $-41$ & $-1075$ & $-3469$ & $-5863$ & $\sim -10^4$ %
\\
$\log z_\textrm{max}$ %
& $8.7$ & $217$ & $698$ & $1180$ & $2465$ %
\\ %
\hline %
\end{tabular}
\label{table:values for n M etc}
\end{table}


\vspace{0.5em}

\textit{Summary and Discussion.} It has been shown in the present
article that it is possible for an oscillating scalar field to
generate repulsive gravity and play the role of dark energy. This
is contrary to the usual thinking that oscillating modes
correspond to excitations or particles that provide nonnegative
pressure and attractive gravity.

The case of the power-law potential is particularly intriguing. In
this case the equation of state of the oscillating quintessence
$w_\textsc{oq}$ is a constant determined simply by the power $n$
in the potential. In particular, when $0<n<1$,
$-1<w_\textsc{oq}<-1/3$ and accordingly the oscillating
quintessence has the ability to provide repulsive gravity and
drive the cosmic acceleration. In addition, the cases
$(n,w_\textsc{oq}) = (1,-1/3),(2,0),(4,1/3)$ are also essential
and interesting.

Several constraints on the parameters in the potential --- the
power $n$, the energy scale $M$ and the current amplitude of the
oscillation $|\phi|_\textrm{max}(0)$ --- are given by observations
and the condition for oscillation. The observational results about
the equation of state and the energy density of dark energy,
$w_\textrm{x} < -0.9$ (1$\sigma$) and $\Omega_\textrm{x} \cong
0.75$, respectively requires $0<n<0.1$ and $M^{4-n}
|\phi|_\textrm{max}^n(0) \sim 10^{-11}\,$eV$^4$. The condition for
oscillation, $HT \ll 1$ from the present back to some early time,
further requires $|\phi|_\textrm{max}(0) \lsim M_\textrm{pl}$ and
gives an upper limit to the redshift, $z_\textrm{max}$, with
respect to the beginning of the oscillation.

Among the examples satisfying the observational constraints and
the oscillation condition, the case where $n \simeq 0.1$
($w_\textsc{oq} \simeq -0.9$) and $M \simeq |\phi|_\textrm{max}(0)
\simeq \rho_\textrm{c}^{1/4} \sim 10^{-3}\,\textrm{eV}$ is
particularly interesting. In this case the only two free
dimensionful parameters in this model, the energy scale $M$ in the
power-law potential and the current amplitude of the oscillation
$|\phi|_\textrm{max}(0)$, are both on the scale of the energy
density of the present universe.
In addition, the period of the oscillation at the present time is
also roughly around the same scale, $\rho_\textrm{c}^{-1/4}$. %
This is much less fine tuned and therefore much more natural,
especially when compared with the case of $V = \frac{1}{2} m^2
\phi^2$ where it is required that $m \lsim H_0 \sim
10^{-33}\textrm{eV}$ (extremely small) and $|\phi| \gsim
M_\textrm{pl} \sim 10^{28}\textrm{eV}$ (extremely large) for
realizing dark energy and driving the cosmic acceleration.

The oscillating quintessence opens a new scenario that is very
different from the running away quintessence
and other slowly rolling evolution patterns involved in most of
the current quintessence models. In this scenario the quintessence
in the recent epoch is oscillating in time, but not slowly rolling
and not always potential-dominated. This oscillating scenario may
be extended to phantom and other dark energy models involving one
or more scalar fields (or even other kinds of fields). This new
scenario largely extends the scope of the model construction for
dark energy played by quintessence and other fields, in
particular, while the oscillation-like behavior (including
orbiting around) is much more familiar and much more frequent to
see in nature of all scales from the very small to the very large.
Accordingly this may provide a link between the microscopic
fundamental physics and the cosmic-scale accelerated expansion
driven by dark energy.


\begin{acknowledgments}
This work is supported by the Taiwan National Science Council (NSC
96-2119-M-007-001).
\end{acknowledgments}

\end{document}